\documentclass[%
showkeys,
 amsmath,amssymb,
 aps,
prb,
]{revtex4-2}
\usepackage{graphicx}
\usepackage{dcolumn}
\usepackage{bm}
\usepackage[T1]{fontenc}
\usepackage[utf8]{inputenc}
\usepackage{amsmath}
\usepackage{braket}
\usepackage{xcolor}

\begin{document}
\title{Superexchange coupling of donor qubits in silicon}

\author{Mushita M. Munia}
\author{Serajum Monir}%
\author{Edyta N. Osika}
\author{Michelle Y. Simmons}
\author{Rajib Rahman}

\affiliation{
 School of Physics, University of New South Wales, Sydney, NSW 2052, Australia.
}%
\affiliation{
 Silicon Quantum Computing Pty Ltd., Level 2, Newton Building, UNSW Sydney, Kensington, NSW 2052, Australia 
}

\begin{abstract}
Atomic engineering in a solid-state material has the potential to functionalize the host with novel phenomena. STM-based lithographic techniques have enabled the placement of individual phosphorus atoms at selective lattice sites of silicon with atomic precision. Here, we show that by placing four phosphorus donors spaced 10-15 nm apart from their neighbours in a linear chain, it is possible to realize coherent spin coupling between the end dopants of the chain, analogous to the superexchange interaction in magnetic materials. Since phosphorus atoms are a promising building block of a silicon quantum computer, this enables spin coupling between their bound electrons beyond nearest neighbours, allowing the qubits to be spaced out by 30-45 nm. The added flexibility in architecture brought about by this long-range coupling not only reduces gate densities but can also reduce correlated noise between qubits from local noise sources that are detrimental to error correction codes. We base our calculations on a full configuration interaction technique in the atomistic tight-binding basis, solving the 4-electron problem exactly, over a domain of a million silicon atoms. Our calculations show that superexchange can be tuned electrically through gate voltages where it is less sensitive to charge noise and donor placement errors. 
\end{abstract}

\keywords{superexchange, silicon qubits, NEMO, FCI, donors, STM}
\maketitle

\section{Introduction}
Donor qubits in silicon are promising candidates for encoding quantum information in the solid state due to their long coherence times \cite{muhonen_storing_2014,stegerQuantumInformationStorage2012,tyryshkin_electron_2012} and their technological link to the silicon platform of the electronics industry. Experimental advancements in the past decades have enabled the precision placement of phosphorus donors in silicon \cite{schofieldAtomicallyPrecisePlacement2003,clarkProgressSiliconbasedQuantum2003,simmonsScanningProbeMicroscopy2005,ruessRealizationAtomicallyControlled2007,fuechsleSingleatomTransistor2012b}. The platform of phosphorus donor based quantum computing has been bolstered by key milestone achievements over the last decade, including single-shot spin-readout \cite{morelloSingleshotReadoutElectron2010}, the realization of single electron and nuclear spin qubits \cite{plaSingleatomElectronSpin2012,pla_high-fidelity_2013}, and more recently, two-qubit SWAP gates
\cite{heTwoqubitGatePhosphorus2019} and three-qubit donor quantum processor with universal logic operation \cite{madzikPrecisionTomographyThreequbit2022}. 
Exchange coupling between the electronic spins of donors remains a key mechanism for fast coupling of two qubits \cite{heTwoqubitGatePhosphorus2019,kaneSiliconbasedNuclearSpin1998}. The exchange interaction depends on the overlap between the electronic wavefunctions and ultimately limits the separation of donor qubits to about 10-15 nm in silicon devices. Long-range coupling schemes through resonators and cavities have recently been explored in donor qubits \cite{tosi_silicon_2017,morse_photonic_2017,Osika2022}, however, these typically require additional fabrication and integration steps adding complexity to the overall manufacturing process. Spacing out the qubits is beneficial from an architectural point of view in fault-tolerant quantum computing \cite{vandersypen_interfacing_2017} as correlations between the qubits due to local noise sources can be minimised. An increase in separation also relaxes stringent gate density requirements and offers more independent electrostatic control of the qubits by reducing their capacitive cross-talk. For STM-patterned donors with phosphorus doped in-plane gates, the density is already low \cite{kiczynski_engineering_2022}, so this technique is particularly appealing. 

In this work, we study long-range exchange coupling between the end spins of four single donor (1P) quantum dots in a linear chain. With each donor containing a single electron, a superexchange coupling is found to emerge between the donors at the end of the chain. This third nearest neighbour interaction enables the qubits to be separated by 30-45 nm. Using atomistic full configuration interaction calculations, we study the eigenvalues and eigenvectors of four electron spins across four 1P atoms in silicon. We investigate the regime of superexchange where the distant qubits can be coherently manipulated and provide guidelines on the appropriate donor placement to achieve this. We also investigate the role of the conduction band valleys on superexchange for donor separation along different crystallographic directions. We simulate the system using realistic electrostatic potentials produced by the surrounding in-plane STM-patterned gates where we demonstrate tunability of superexchange with gate voltages, a crucial requirement for the realization of electrically-controlled singlet-triplet oscillations. Finally, we comment on the sensitivity of superexchange to charge noise and donor placement errors, as well as the role of nuclear spins in singlet-triplet oscillations induced by superexchange.

\begin{figure*}[htb!]
    \centering
    \includegraphics[width=\textwidth]{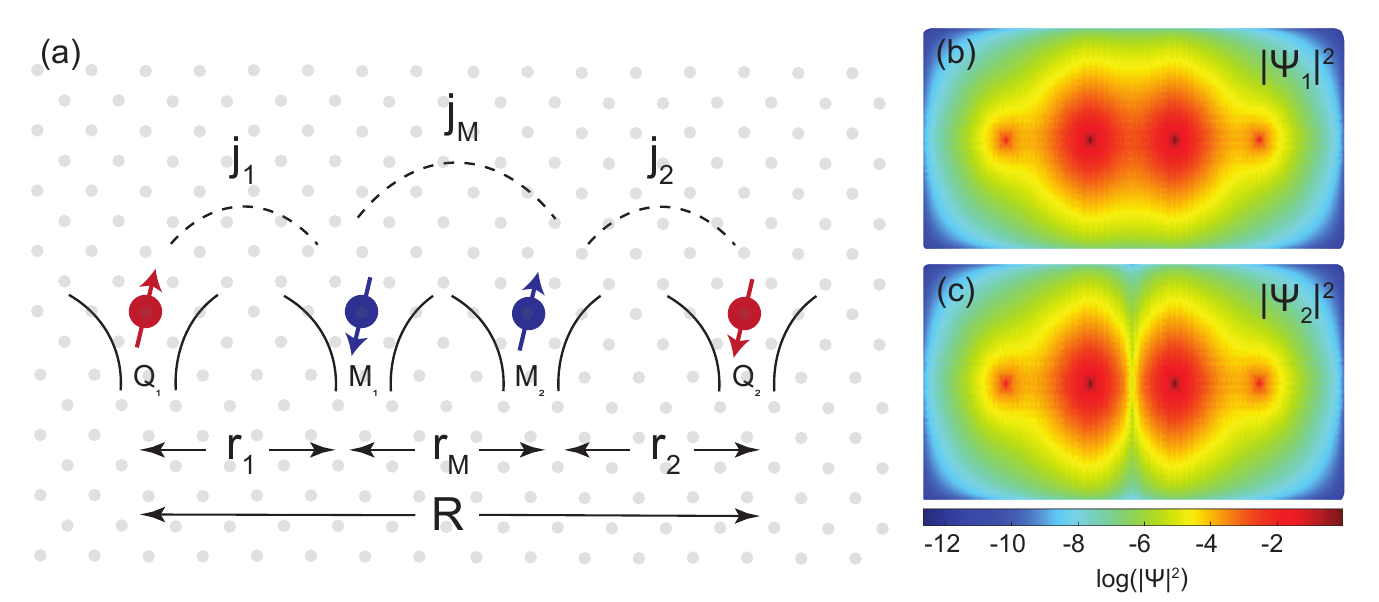}
    \caption{\textbf{Schematic representation of a simulation using four single phosphorus donors each with an electron. (a)} Four phosphorus donors placed in the silicon crystal (grey) where the electrons localized in the two middle donors (blue) are mediators ($M_1$ and $M_2$) for the electron spins localized in the end donors ($Q_1$ and $Q_2$) (red). The separation of the blue mediator spins is $r_M$ and their exchange coupling is $j_M$. The first qubit $Q_1$ is separated from the first mediator $M_1$ by $r_1$ ($\sim$ 10 nm) and the second qubit $Q_2$ is separated from the second mediator $M_2$ by $r_2$ ($\sim$ 10 nm). The corresponding exchange coupling between them is $j_1$ and $j_2$ respectively. The total separation of the qubits $Q_1$ and $Q_2$ is $R$ ($\sim$ 30 nm). 
    The probability density of the lowest single electron valley-orbital \textbf{(b)} bonding and \textbf{(c)} anti-bonding state in a logarithmic scale calculated using an atomistic tight-binding method.}
    \label{fig:schematics}
\end{figure*}

Recent experiments and theoretical calculations have shown indirect coupling of distant electrons in quantum dots via a central empty mediator \cite{chan_exchange_2021,deng_interplay_2020, rancic_ultracoherent_2017, baart_coherent_2017,mehl_two-qubit_2014}, multi-electron quantum dot \cite{deng_interplay_2020,mehl_two-qubit_2014, malinowski_fast_2019,malinowski_spin_2018,deng_negative_2018,croot_device_2018,srinivasa_tunable_2015} and linear chain of singly-occupied quantum dots \cite{chan_exchange_2021,qiao_long-distance_2021,fei_mediated_2012}. However, to date, comprehensive studies have not been performed on long-range indirect coupling for donor qubits in silicon. Compared to electrostatically defined quantum dots, donor quantum dots in silicon typically have atomic-scale properties with wavefunction length scales an order of magnitude less and large quantity of orbital-valley energy splittings \cite{ramdas_spectroscopy_1981}. A single phosphorus donor can bind at most one or two electrons with the nucleus having a net $1/2$ spin. Early papers on single donor qubits showed how the exact position and axis of separation of donor qubits can significantly affect their direct exchange, including such effects as exchange oscillations due to valley interference \cite{tankasalaShallowDopantPairs2022b, voisinValleyInterferenceSpin2020}. More recent work has shown that these effects can be mitigated by separating the donors along specific crystallographic directions, by using multi-donor quantum dots \cite{wangHighlyTunableExchange2016a} or by applying strain or placing the donors close to the surface \cite{tankasalaShallowDopantPairs2022b}. Highly tunable nearest-neighbour exchange has also been predicted \cite{wangHighlyTunableExchange2016a} and demonstrated \cite{heTwoqubitGatePhosphorus2019} in asymmetric donor quantum dots, showing that multi-electron molecular physics can be tuned by gate voltages. However, the atomistic character of these donors and donor dot systems needs to be accounted for when considering indirect couplings like superexchange since the phenomenon emerges from individual nearest-neighbour exchange couplings.

\section{Methods}
The exact calculation of a multi-electron system is challenging because of the complicated and numerically intensive electron-electron interaction term in the Schr\"odinger equation \cite{szaboModernQuantumChemistry1996}. The accuracy relies both on the quality of the basis states and the multi-electron approximation. In semiconductor materials such as GaAs, eigenstates from effective Hubbard Hamiltonian \cite{chan_exchange_2021,malinowski_fast_2019,chan_sign_2022} or simple Fock-Darwin states are often used as basis states \cite{deng_interplay_2020}. These are however unsuitable for silicon due to the multi-valley states. The effective mass approximation has also been used for calculating single electron basis states for donors in silicon \cite{saraivaTheoryOneTwo2015,joeckerFullConfigurationInteraction2021}. However, the theory does not provide a complete description of the band structure and can be inaccurate at higher energy levels. On the contrary, the full-band atomistic tight-binding model has the potential to capture all intricate nuances of the wavefunction of phosphorus donors in silicon hence its use in this work. 

The most common approximation for multi-electron calculations is the Hartree-Fock method which neglects the electron-electron correlations to minimize the complexity of the calculations. Configuration interaction (CI), on the other hand, enhances the accuracy in the treatment of many-body interactions but can be computationally intensive. This is particularly true for a multi-valley material like silicon where only a few electron calculations are present in the literature \cite{tankasalaShallowDopantPairs2022b,tankasalaTwoelectronStatesGroupV2018}. In this paper, we have performed state-of-the-art investigation combining atomistic basis states with a full configuration interaction method to calculate the energy levels of a four-electron donor system without compromising accuracy. 

\subsection{Atomistic Full Configuration Interaction}
 Full configuration interaction (FCI) is a method to obtain the exact numerical solution of a many-body Schr\"odinger equation, limited by the number and quality of single electron basis states. The single electron basis states used here are calculated using a 10-band $sp^3d^5s^*$ atomistic tight-binding (TB) method in NEMO3D \cite{ahmedMultimillionAtomSimulations2009,klimeckAtomisticSimulationRealistically2007a}. This approach uses a localized atomic orbital-based method with nearest-neighbour interactions. The TB parameters are optimized to reproduce the bulk silicon band structure \cite{boykinValenceBandEffectivemass2004}. The phosphorus donors are represented using a Coulomb potential with a central cell correction at the donor site which can successfully determine the experimentally measured energy spectra of donors in silicon \cite{rahmanHighPrecisionQuantum2007,rahmanOrbitalStarkEffect2009}.

 The schematic representation of the simulations is illustrated in Figure \ref{fig:schematics}(a). There are four phosphorus donors placed in a chain, each with an electron. The end electron spins, $Q_1$ and $Q_2$ are the qubits separated by R. The middle spins, $M_1$ and $M_2$ work as mediators with corresponding donors being separated by $r_M$. The separation between $Q_1$ and $M_1$ is $r_1$ and the separation between $M_2$ and $Q_2$ is $r_2$. The simulation domains used in the calculations entail $\sim1.14$ million atoms which account for $\sim60$ nm of silicon in the direction of separation and $\sim20$ nm in the other two directions. The single electron basis states, solved from a parallel Block Lanczos eigensolver, are used to calculate the anti-symmetric Slater determinants of the multi-electron problem. The lowest two single electron valley-orbital states of the system are shown in Figure \ref{fig:schematics}(b) and (c). Here we see bonding and anti-bonding state formation in the middle two donors $M_1$ and $M_2$ of a four-donor chain. The rest of the single electron molecular basis states are shown in Figure \ref{fig:basis_states} in the Supplementary Information where we see the next two valley-orbital states are mainly localized in the outer donor dots.

 The four-electron wavefunction is a superposition of various symmetry-permitted configurations of the Slater determinants. All possible integrals between the Slater Determinants with pairwise electronic interaction operators are computed to capture Coulomb, exchange, and higher-order correlations. The four electron Hamiltonian constructed from the Slater determinants is solved using the block Krylov-Schur algorithm within the Trilinos framework \cite{bakerAnasaziSoftwareNumerical2009}. The number of single electron basis states in FCI calculations is increased until the eigenvalues of the four-electron system converge within a chosen tolerance -- see Supplementary Information. On average, 56 single-electron basis states are sufficient to reach convergence. The eigenvalue of the ground state is separated from the triply degenerate excited states by the indirect exchange coupling. We place the donor atoms in different separations and orientations in our simulations and calculate this exchange coupling. 
 
\subsection{Effective spin Hamiltonian}
To analyze and interpret our FCI results, we compare them with an effective Hamiltonian model \cite{qiao_long-distance_2021}. We can represent the four-spin system with a spin Hamiltonian such as - 
\begin{equation}
    H_{eff} = \frac{j_1}{4} \sigma_1 . \sigma_2 + \frac{j_M}{4} \sigma_2 . \sigma_3 + \frac{j_2}{4} \sigma_3 . \sigma_4 
    \label{Heff}
\end{equation}
where $\sigma_i$ is a Pauli matrix corresponding to an electron spin located on $i^{th}$ donor in the chain. $j_M$ is the exchange coupling between the middle two spins, $M_1$ and $M_2$. $j_1$ is the exchange coupling between $Q_1$ and $M_1$ and $j_2$ is the exchange coupling between $Q_2$ and $M_2$ -- see Figure \ref{fig:schematics}(a) for the schematic of the system. 

We only consider the subspace with spin-zero states ($S_z = 0$) to construct the Hamiltonian as we are interested in the singlet-triplet oscillations in the qubits $Q_1$ and $Q_2$. When $j_1,j_2 \ll j_M$, we can isolate the low energy states of the four-electron spin Hamiltonian using a Schrieffer-Wolff transformation. In that case, the effective Hamiltonian in a Heisenberg exchange form, $H_{SW}$ \cite{qiao_long-distance_2021} now becomes - 
\begin{equation}
\begin{split}
    H_{SW} & = \frac{J_{SW}}{4} \sigma_1 . \sigma_4 \\
    J_{SW} & = \frac{j_1j_2}{2j_M} [1+\frac{3(j_1+j_2)}{4j_M}]
\end{split}
\label{HSW}
\end{equation}

In the regime where $j_1, j_2 \ll  j_M$, the lowest energy eigenstates are characterized by the singlet state formed within the two middle dots and singlet or triplet states formed within the outer dots \cite{qiao_long-distance_2021}. The energy difference, $\Delta E$ between the lowest long-distance singlet and triplet states, i.e.
\begin{equation}
\begin{split}
    S^l \approx 1/\sqrt{2}(\ket{\uparrow S \downarrow} -  \ket{\downarrow S \uparrow}) \\
    T^l_0 \approx 1/\sqrt{2}(\ket{\uparrow S \downarrow} +  \ket{\downarrow S \uparrow})
\end{split}
\end{equation}
is what we call superexchange ($\Delta E$) in this paper -- see Figure \ref{fig2}(a) for the energy level diagram of the system. For $j_1, j_2 \ll  j_M$ the middle-dot singlet manifold is well separated from all the middle-dot triplet states desirable for coherent coupling of the outer spins only, without any interference from the middle-dot spins.

\section{Results and Discussions}
\subsection{Equal nearest-neighbour separation of donors}

\begin{figure*}[tb!]
    \centering
    \includegraphics[width=\textwidth]{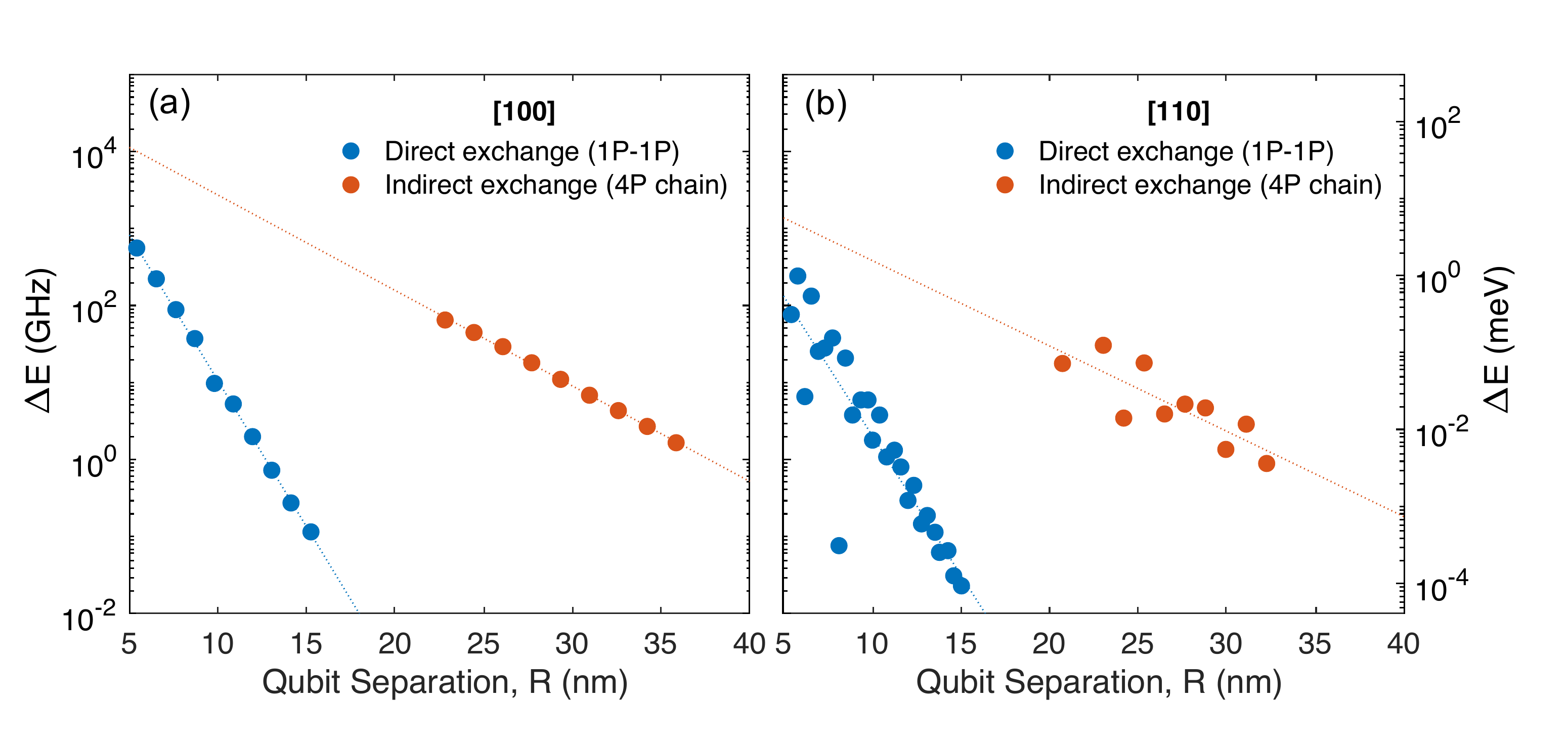}
    \caption{\textbf{Comparison of indirect and direct exchange coupling of 1P donors in silicon calculated using atomistic FCI (a)} Comparison of indirect exchange (4P chain) of the end-spins (red dots) with nearest-neighbour exchange coupling (1P-1P) (blue dots, replicated from \cite{wangHighlyTunableExchange2016a}) along the $[100]$ direction. Dashed lines provide linear fits to both data sets. Here we see that the indirect exchange is higher than the nearest-neighbour exchange, with the exponential dependence with separation being less steep for indirect coupling. \textbf{(b)} Same as (a) but along the $[110]$ direction. Here we see a similar trend in terms of the comparison between direct and indirect exchange as for the $[100]$ direction. However, we also observe oscillations in the exchange coupling along this crystalline orientation due to valley quantum interference.}
    \label{fig1}
\end{figure*}

We first analyze the equidistant case, where the separation between each neighbouring pair of donors is equal, $r_1=r_2=r_M=R/3$ -- see Figure \ref{fig:schematics}. Here we have approximately equal values of exchange coupling between each pair of donors, i.e. $j_1 \approx j_2 \approx j_M$. In Figure \ref{fig1} (with red dots), we show the values of the indirect exchange coupling for these equispaced donors (4P chain) with constant but increasing separation between the dots along the $[100]$ (left) and $[110]$ directions (right) in the silicon crystal, calculated using FCI. For comparison, we also plot (with blue dots) the direct exchange values for two donors, i.e. 1P-1P system, separated by R, as previously calculated in \cite{wangHighlyTunableExchange2016a}. Looking at the fitted dashed lines between the 4P chain and 1P-1P cases, we can see that the presence of the mediator donors dramatically increases the exchange coupling between the two outer spins. The indirect exchange coupling for a donor chain of $R$ of about 25 nm reaches $\sim10^{2}$ GHz ($10^{-1}$ meV), while for direct exchange for the same separation without the presence of mediators would fall below $10^{-4}$ GHz ($10^{-7}$ meV) (extrapolated from the dataset in \cite{wangHighlyTunableExchange2016a}). 

We also see that the direct exchange decays much faster than the indirect exchange by comparing the slope of these two plots, true for both [100] and [110] crystal directions. For the [100] case in Figure \ref{fig1}(a), the slope of the fitted dashed line for the direct exchange as a function of qubit separation is $\sim-0.38$/nm  whereas the slope for the indirect exchange is $\sim-0.12$/nm. Similarly in the [110] case in Figure \ref{fig1}(b), the slope of the fitted line for the direct exchange is $\sim-0.36$/nm whereas the slope for the indirect exchange is $\sim-0.11$/nm. In general, the indirect exchange decays almost three times less with donor separation than the direct exchange. This dependency with distance lets the qubits be far separated in the device and allows less correlated noise between them while maintaining strong spin coupling. Since the nearest neighbour exchange coupling has an exponential dependence with increasing separation of the donors, one might expect superexchange to change as an exponential function of $r_1+r_2-r_M$ -- see Equation \ref{HSW}. On the contrary, the nearest neighbour exchange coupling changes as an exponential function of the qubit separation $R=r_1+r_2+r_M$ which results in a faster decay of direct exchange than the indirect one. 

We observe no oscillations in the exchange coupling when the donors are separated in the [100] direction whilst we see oscillations in the exchange energy for separation in the [110] orientation, a signature of inter-valley quantum interference arising from the band structure of silicon \cite{tankasalaShallowDopantPairs2022b,wangHighlyTunableExchange2016a,koillerExchangeSiliconbasedQuantum2001}.

\begin{figure}[htb!]
    \centering
    \includegraphics[width=\textwidth]{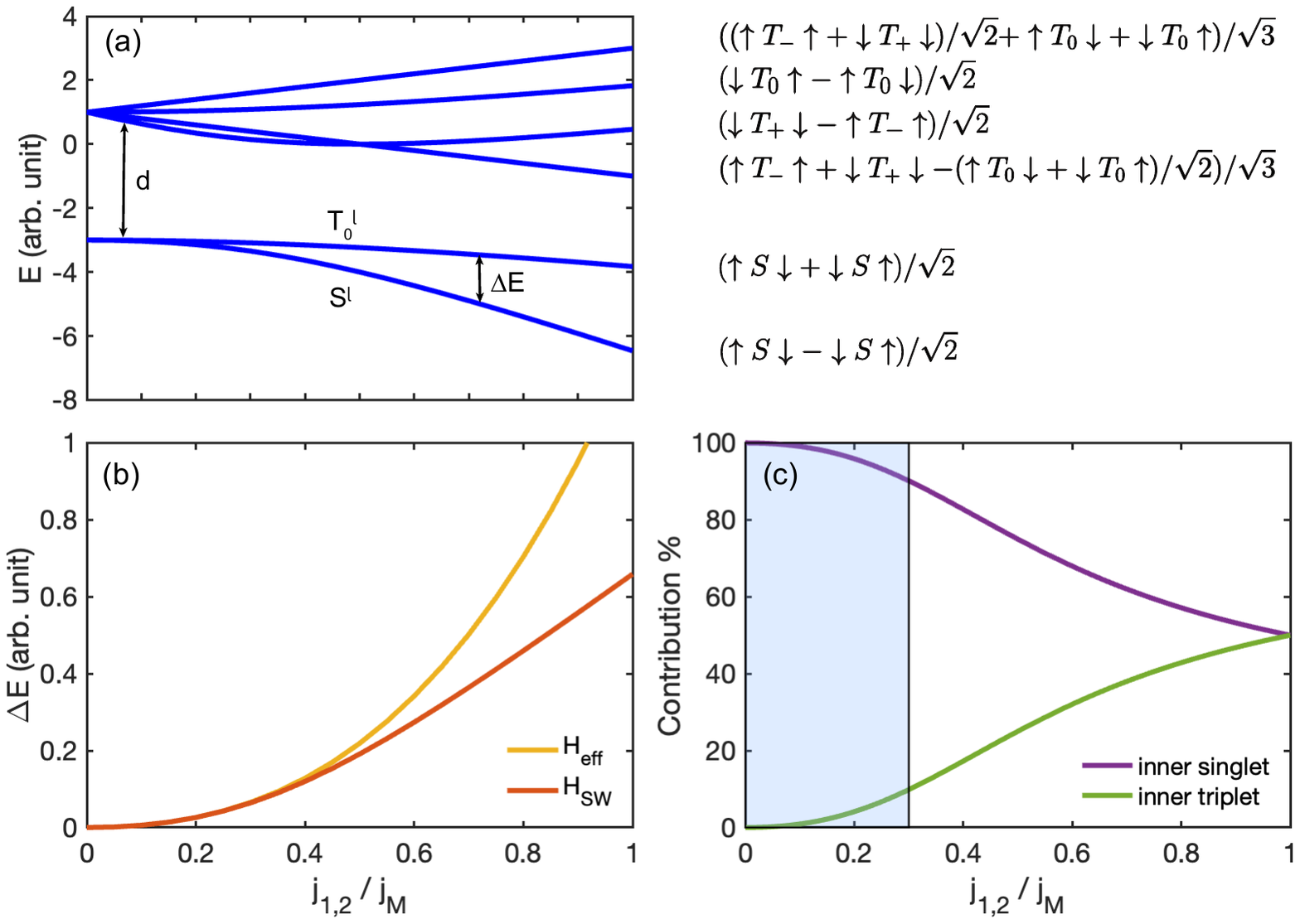}
    \caption{\textbf{Estimate of exchange coupling strength from the spin Hamiltonian. (a)} Energies of 4-electron eigenstates calculated with the effective Hamiltonian $H_{eff}$ as a function of $j_{1,2}/j_M$. The two lowest states (spanned by $\ket{\uparrow S \downarrow}$ and $\ket{\downarrow S \uparrow}$) are well separated from the higher states when $j_{1,2}/j_M$ is smaller. The separation of these two energy levels is the superexchange $\Delta E$. The labels on the right refer to the dominant contribution of the corresponding energy states when $j_{1,2}/j_M$ $\sim$ $0$. \textbf{(b)} A comparison of the exchange coupling, $\Delta E$ calculated from the effective spin and the Schrieffer-Wolff Hamiltonian shows that up to approximately $j_{1,2}/J_M$ $=$ $0.3$, the superexchange from $H_{eff}$ and $H_{SW}$ is the same since the Schrieffer-Wolff Hamiltonian is a valid approximation of the spin Hamiltonian. Beyond this regime, their behaviour diverges significantly because the Schrieffer-Wolff transformation no longer holds indicating the development of an admixture in the singlet-like state of the middle two spins. \textbf{(c)} Contribution of the inner-singlet and inner-triplet basis states to the ground state of $H_{eff}$ as a function of $j_{1,2}/j_M$. Here we see that, as $j_{1,2}/j_M$ increases, the singlet contribution from the inner mediator spins decreases, and the triplet contribution increases. At $j_{1,2}/j_M$ $=$ $1$, the ground state has equal contributions from singlet and triplet mediator spin states. The contribution of the inner singlet to the ground state is more than 90\% for $j_{1,2}/J_M<0.3$ (shaded region).}
    \label{fig2}
\end{figure}

\subsection{Different regimes of superexchange from effective spin Hamiltonian}
To obtain superexchange and coherent control between the electron spin qubits at the end of the chains $Q_1$ and $Q_2$, it is essential that the middle spins $M_1$ and $M_2$ form a singlet-like state. Otherwise, there will be additional oscillations originating from the middle spins impeding coherent manipulation of the qubits. Thus we are looking for an operational regime where the ground state and the first excited state are well separated from the higher energy states in which the indirect coupling can be termed superexchange. In Figure \ref{fig2}(a) we plot the energies of all the eigenstates of the effective Hamiltonian $H_{eff}$ as a function of $j_{1,2}/j_M$ varying from 0 to 1. When the outer spins are weakly coupled with the middle spins ($j_{1,2}/j_M\sim0$), we see two clear branches of energy states separated by $d$ but the separation between the branches decreases as $j_{1,2}/j_M \sim 1$. 

In Figure \ref{fig2}(b), we plot the energy difference between the two lowest states, as a function of $j_{1,2}/j_M$ as calculated from $H_{eff}$ and $H_{SW}$. These two lowest states are the long-distance singlet and triplet states, namely $S^l$ and $T^l_0$. Here we see that, in the regime where $j_{1,2}/j_M\sim0$, the solutions from the effective spin and the Schrieffer-Wolff Hamiltonian are the same since the assumption of the transformation is valid here. As $j_{1,2}/j_M$ increases, the two solutions start to diverge suggesting the Schrieffer-Wolff transformation no longer holds. The lowest two energy states (spanned by $\ket{\uparrow S \downarrow}$ and $\ket{\downarrow S \uparrow}$) are no longer well separated from the excited states and there are now triplet-like admixtures in the singlet-like state of the middle two spins. 

In Figure \ref{fig2}(c), we can see the contributions in the ground state $S^l$ of the inner-dot singlet states (i.e. $\ket{\uparrow S \downarrow}$ and $\ket{\downarrow S \uparrow}$) and inner-dot triplet states (all the remaining basis states). When $j_{1,2}/j_M=1$ (i.e. the equidistant case where all donors are equally separated), the inner-dot singlet and triplet contributions in $S_l$ are both approximately 50\% and we no longer have a well-defined two-level system of $S^l$ and $T^l_0$, but now have contributions from the four higher states shown in \ref{fig2}(a). As $j_{1,2}/j_M$ decreases, the contribution of the inner-dot singlet reaches more than $90\%$ where $j_{1,2} <\sim0.3 j_M$ (shaded region). This regime can be considered as a threshold for coherent operation of the qubits. However, we note that the threshold we mention here ($j_{1,2}/j_M\sim0.3$) is not the upper limit of the coherent regime. The ratio should ideally be very close to $0$.

\begin{figure*}[htb!]
    \centering
    \includegraphics[width=\textwidth]{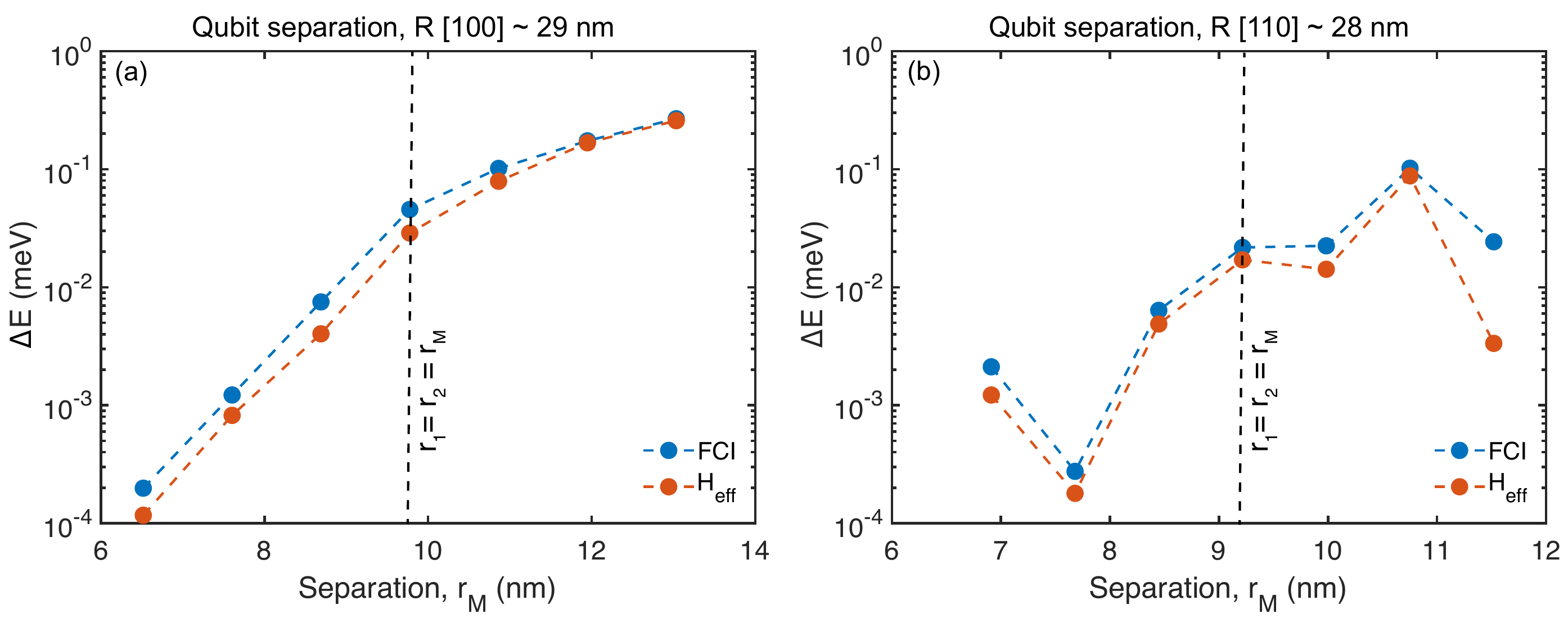}
    \caption{\textbf{Exchange coupling as a function of different middle donor separations, $r_M$.} The exchange coupling $\Delta E$ is calculated using atomistic FCI for different separations of the middle donors $r_M$ in the (a) [100] and (b) [110] direction while keeping the total outer qubit separation fixed (R = 29.3274 nm for [100] and R = 27.648 nm for [110]). The system is symmetric ($r_1$ $=$ $r_2$) to the end donors. The results show that the dependence of exchange energy changes on either side of the equidistant separation (dashed grey line) for all donors. The dashed line indicates where $r_1$ = $r_2$ = $r_M$ and the left region of this line where $r_1
    $ $=$ $r_2$ $>$ $r_M$ is where the two middle donors form a singlet state and the exchange coupling between the end donors is termed superexchange. We compare our FCI results with the effective spin Hamiltonian, $H_{eff}$ of Eq. \ref{Heff} with $j_{1,2}$ and $j_M$ extracted from \cite{wangHighlyTunableExchange2016a}. For both cases, the results show reasonable agreement with each other. We attribute any differences mainly to the fact that the total confinement potential of 4 donors is deeper than that of 2 donors which results in a slightly higher exchange coupling $\Delta E$ in our atomistic FCI calculations.}
    \label{fig3}
\end{figure*}

\subsection{Modulating superexchange by changing the middle donor separation}
To explore the coherent-control regime of the 4-donor chain using our FCI calculations, we switch from the equidistant case discussed in Figure \ref{fig1} to a chain with varied $r_{1,2}/r_M$ ratios, see Figure \ref{fig3}. Here, we present values of the indirect exchange coupling for chains oriented along the [100] and [110] direction in \ref{fig3} (a), (b) respectively, as a function of middle donor separation while keeping the outer donor separation $R$ constant. The lower middle donor separation corresponds to a higher exchange coupling $j_M$ and lower $j_1$ and $j_2$ couplings, moving the system closer to the regime well described by the Schrieffer-Wolff approximation. In general, we see from Figure \ref{fig3} that the exchange coupling exponentially decreases as we decrease the middle donor separation. The grey dashed line represents the separation where $r_1=r_2=r_M$. The exchange coupling shows two different dependencies on either side of this point. The first is where the distance between the middle donors is smaller than the values of $r_1$ or $r_2$ and the second is where it is large. For donors separated along the [110] direction, we see clear oscillations in $\Delta E$ as a function of $r_M$.

As we have discussed in the previous section the equidistant donor chain is not suitable for coherent manipulation of distant spins due to the admixture of states. The right side of the grey dashed line where $r_1=r_2 \leq r_M$ is therefore not valid for operating qubits. To specify the limits of these different regimes, we use the values of direct exchange from the 1P-1P results in Ref. \cite{wangHighlyTunableExchange2016a} to find $r_{1,2}$ and $r_M$ and the corresponding threshold. Decreasing the separation of the middle dots gives rise to a value of the superexchange that decreases exponentially while increasing the inner singlet contributions to the ground state -- see Table 1 in the supplementary information. So there is a trade-off between the strength of coupling and operation fidelity. However, we see even small changes in the middle donor separation, of the order of $\sim$ 2 nm, can increase the inner singlet contributions to the ground state from $\sim49\%$ to $\sim98\%$ while keeping the value of superexchange still high enough ($\sim100$ MHz or $\sim1\times10^{-3}$ meV) for use in realistic devices.

\subsection{Electrical control of superexchange}
\begin{figure}[htb!]
    \centering
    \includegraphics[width=\textwidth]{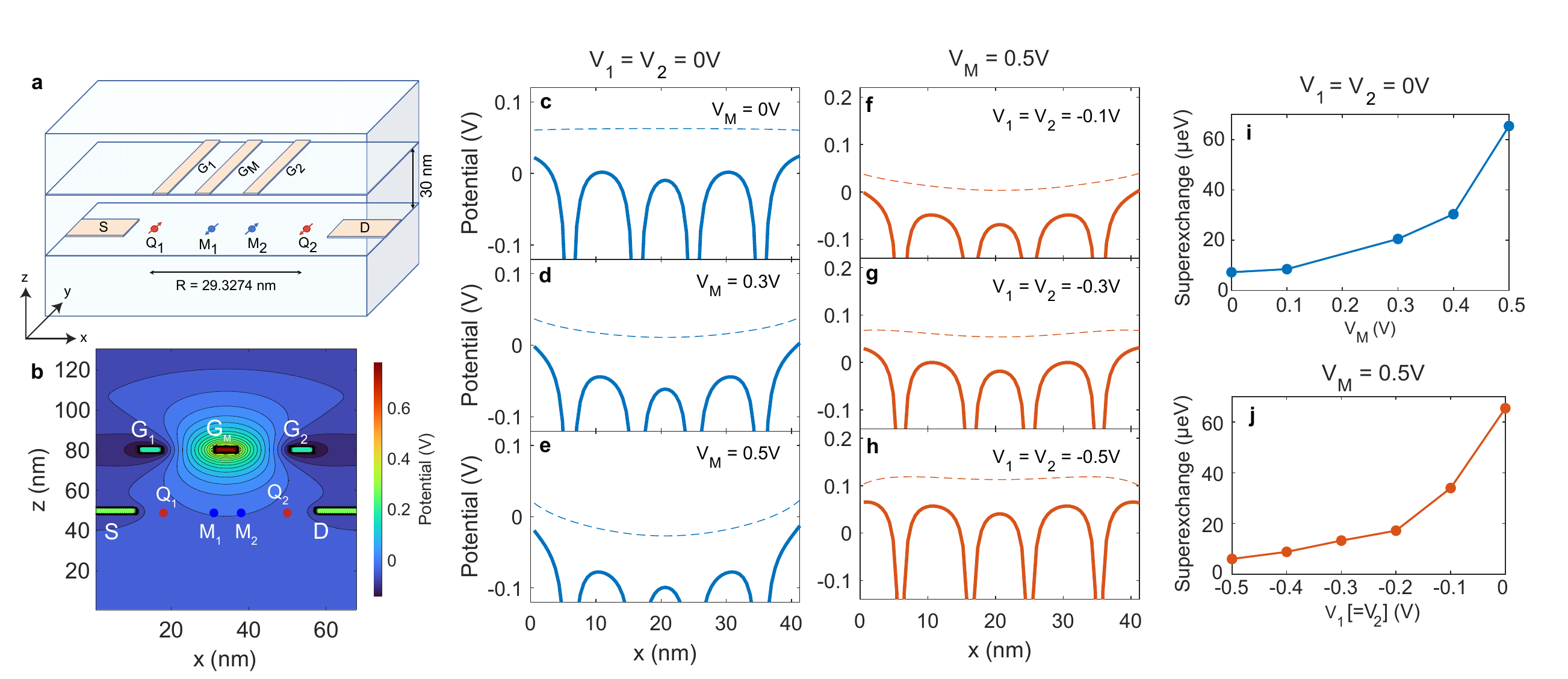}
    \caption{\textbf{Modulating superexchange using voltage biases on phosphorus-doped top gates above the donor plane.} Electrostatic potential of the device in the \textbf{(a)} Schematic representation of the simulation domain. The qubit donor atoms ($Q_1$ and $Q_2$) and the mediator donor atoms ($M_1$ and $M_2$) are placed in-plane between the source and drain (S, D) electrodes. Three phosphorus-doped top gates ($G_1$, $G_2$, and $G_M$) are placed 30 nm above the donor plane. \textbf{(b)} The total potential of the device in the x-z plane (cut taken at donor location). Here the voltage of gates $G_1$ and $G_2$ are $V_1$ = $V_2$ = $-0.1V$ whilst the middle gate $G_M$ has a voltage $V_M$ = $0.5V$ and S and D are grounded. The red and blue dots represent the location of the donors and are not to scale. 1D potentials with cuts taken at donor location when $V_1$ = $V_2$ = 0V and \textbf{(c)} $V_M$ = 0V, \textbf{(d)} $V_M$ = 0.3V and \textbf{(e)} $V_M$ = 0.5V respectively as we deplete the barrier between the donors. The dashed lines represent the potential from the gates and the solid lines represent the individual donor potentials combined with the electrostatic potential from the gates. As the voltage of the top gate increases, the curvature of the potential also increases. Same as (c-e) when $V_M = 0.5V$ and \textbf{(f)} $V_1 = V_2$ = -0.1V, \textbf{(g)} $V_1 = V_2$ = -0.3V and \textbf{(h)} $V_1 = V_2$ = -0.5V respectively. \textbf{(i)} Superexchange as a function of $V_M$ when $V_1$ = $V_2$ = 0V. We see that the magnitude of the superexchange is increasing as a positive bias is applied to $G_M$. The increased curvature of the applied potential of $V_M$ = $0.5V$ (comparing (c) and (e)) gives rise to the  modulation of the magnitude of superexchange by a factor of 10. \textbf{(j)} Superexchange as a function of $V_1$ = $V_2$ when $V_M$ = $0.5V$. Here we see the superexchange is decreasing as more negative bias is applied as a result of decreased potential curvature (comparing (f) and (h))} 
    \label{fig:electrostatics}
\end{figure}

To address the tunability of superexchange, we have applied an electric field along the direction of the donor chain in our simulations where all donors are separated by $9.7758$ nm ($r_1$ = $r_2$ = $r_M$ = 9.7758 nm). For a realistic applied electric field of $2$ MV/m, we observe an exchange coupling of 18.09 GHz ($74.83\,\mu$eV) compared to 11.01 GHz ($45.556\,\mu$eV) under no electric field. The electric field detunes the qubits and increases the virtual tunnelling between them which results in slightly increased (less than a factor of 2) superexchange. However, when the donors are all charge neutral with one electron on each phosphorus atom, it is difficult to reach the (2,1,1,0) charge regime from a (1,1,1,1) regime without applying a very large bias. A similar challenge has been observed in \cite{wangHighlyTunableExchange2016a} where applying an electric field of 2 MV/m the exchange coupling was modulated by 5 times. In the case of superexchange, the sensitivity of the coupling with an applied electric field is even less, hence, the distanced qubits are less susceptible to charge noise caused by electric field fluctuations compared to the nearest neighbour ones. 
To explore different levels of control over superexchange, we apply a J-gate bias on a system with a \\ phosphorus-doped silicon top gate based on the original Kane architecture \cite{kaneSiliconbasedNuclearSpin1998}. We place gates 30 nm above the donor plane to control the potential barrier between donor atoms and the nearest neighbour exchange and observe their effect on superexchange. Similar three-dimensional tuning of the potential barrier between in-plane phosphorus-doped gates has been recently experimentally demonstrated \cite{donnelly2021monolithic} in precision engineered STM tunnel junctions where the gates were degenerately phosphorus-doped silicon layers. 
The schematic of the system is illustrated in Figure \ref{fig:electrostatics}(a). The device consists of five quasi-metallic gates (source S, drain D, and three top gates - $G_1$, $G_2$ and $G_M$). $G_1$ and $G_2$  are used to tune $j_1$ and $j_2$ and $G_M$ is used to tune $j_M$. The total electrostatic potential profile of the device in the x-z plane is shown in Figure \ref{fig:electrostatics}(b) where a cut is taken at the donor location. Here we see that, in the donor plane, the top gates create a parabolic potential profile. This total electrostatic potential is added to the tight-binding Hamiltonian while solving the energy levels of the donor quantum dots. In Figure \ref{fig:electrostatics}(c-h), 1D cuts are taken at the donor locations shown for different $V_M$ and $V_1$(=$V_2$), i.e. for different voltages applied at gates $G_M$ and $G_1(G_2)$, respectively. Here S and D are grounded, i.e. $V_S = V_D = 0$. We note that just the presence of the electrostatic leads around the donors also modifies the total electrostatic potential of the device, even without an applied voltage. This is due to the band structure mismatch between the leads and the surrounding material, and the band-bending caused by the leads \cite{ryu2013atomistic}. We calculate the total electrostatic potential profile of the device using a multi-scale modelling technique that combines the atomistic calculation of the band structure of the gates with a non-linear Poisson solution of the entire device \cite{donnelly2023multi}. 

\subsubsection{Applying voltage to the middle top gate, $G_M$}
Since superexchange $\Delta E$ is a function of nearest-neighbour exchanges (i.e. $j_1$, $j_2$ and $j_M$), it is necessary to understand how these exchange couplings change under an applied bias. When we apply a positive voltage to the middle gate $G_M$, the potential barrier between the mediators $M_1$ and $M_2$ decreases but $j_M$ would increase - see Figure \ref{fig:electrostatics}(c-e) for increasing values of $V_M$. However, from the dashed line in Figure \ref{fig:electrostatics}(e), we see that the applied bias in $G_M$ detunes the qubits $Q_1$ and $Q_2$ with respect to $M_1$ and $M_2$ so $j_1$ and $j_2$ would also increase. As we can see from Equation \ref{HSW}, the superexchange strongly depends on the interplay between $j_M$ and $j_1, j_2$. In our case, here the superexchange increases overall with $V_M$ since the effect caused by detuning between the qubits and the mediator donors is stronger than from lowering the middle barrier -- see the potential profile in Figure \ref{fig:electrostatics}(e). We observe this effect on the value of $\Delta E$ in Figure \ref{fig:electrostatics}(i), where we can see the superexchange as a function of $V_M$ for $V_1=V_2=0$ (case corresponding to \ref{fig:electrostatics}(c-e)). By changing the middle gate potential from 0 to 0.5V, we modulate the magnitude of the superexchange by a factor of 10. The increase of superexchange is due to a relatively higher increase of $j_1, j_2$ than $j_M$, resulting in lesser singlet contribution from the mediator dots, as discussed in the previous section. In our case, the singlet contribution in the middle dot drops from 92.2\% ($V_M = 0$) to 67\% ($V_M = 0.5V$). 
\subsubsection{Applying voltage to all three top gates, $G_1$, $G_2$ and $G_M$}
We can minimize the detuning of the qubits $Q_1$ and $Q_2$ by applying a negative voltage to the left and right gates. Then the superexchange would decrease back - here from 65.42 $\mu$eV to 6 $\mu$eV when changing $V_1=V_2$ from 0 to -0.5 V for $V_M=0.5$ V -- see Figure \ref{fig:electrostatics}(j). The positive bias applied to $G_M$ increases $j_M$ and the negative bias applied to $G_{1,2}$ decreases $j_{1,2}$, so the $j_{1,2}/j_M$ ratio becomes smaller. The potential profile at the donor location for $V_M = 0.5V$ and $V_1$ = $V_2$ = -0.1V, -0.3V and -0.5V is shown in Figure \ref{fig:electrostatics}(f-h). If we look at the inner singlet contributions on the ground state, we see that the contribution has increased up to 94.05\% when $V_1$ = $V_2$ = -0.5V. Tuning superexchange by using three gates might be useful in increasing the fidelity of the operation, otherwise, the tuning shows a similar range of superexchange even with one top gate.  
With these simulations, we have shown that it is possible to manipulate superexchange in a 4P chain by purely electrostatic means. Applying voltage through top gates exhibits better tunability of superexchange than a tilt voltage along the donor separation. We have shown that it is possible to modulate superexchange by a factor of 10 with modest gate potentials of 0.5 V. We note that even higher tunability might be desired to turn the coupling on and off for qubit operations. Such high tunability of the exchange coupling can also be achieved by depleting the middle donors. Without electrons in the middle donors, the qubits will be coupled by direct exchange which is significantly low at these separations. On the other hand, loading electrons in the middle donors will introduce a high value of superexchange. This mechanism can be crucial for the realization of singlet-triplet oscillations in distanced qubits. 

\subsection{Effect of asymmetric separation of the donors}
Current STM lithography techniques allow us to place donor atoms in silicon with the precision of a single lattice constant \cite{fuechsle2012single,PhysRevApplied.16.054037,bussmann2021atomic,Simmons2022,wyrick_enhanced_2022}. As a consequence, we can investigate how small shifts in the exact P atom location within the four donor chain can affect the value of superexchange available. For that purpose, we analyze a chain oriented along the [100] direction with constant middle dot separation $r_M=5.431$ nm but with variable positions of the outer donors, described by $r_1$ and $r_2$ where $r_1,r_2 \geq$ 7 nm. We have deliberately chosen a smaller value of $r_M$ so that $r_1,r_2 > r_M$. We do not consider nearest neighbour separations less than 7 nm (qubit separation $R<$ 20 nm) since we are interested in the long-distance qubit coupling regime and want to exceed the distance achievable by direct exchange. 

\begin{figure}[htb!]
    \centering
    \includegraphics[width=10cm]{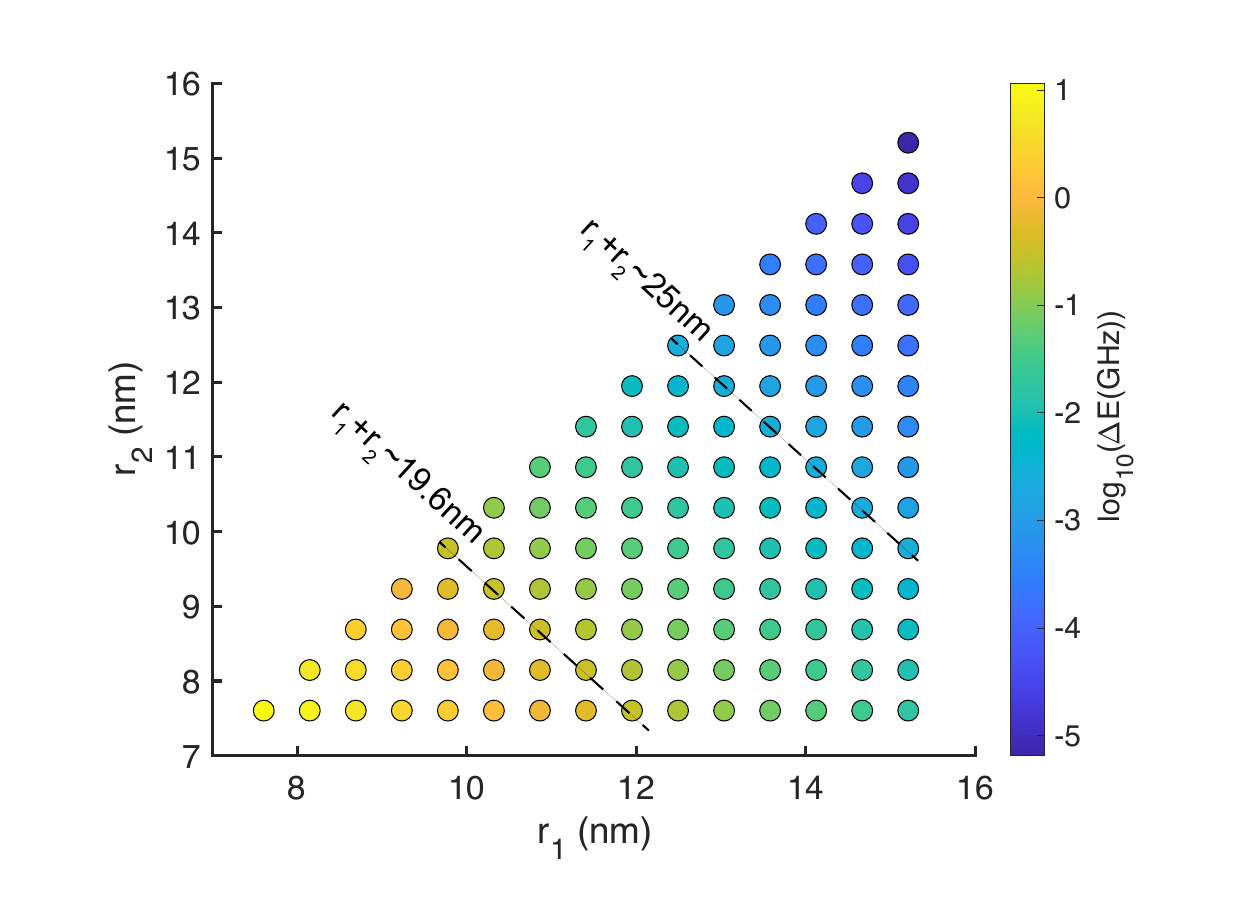}
    \caption{\textbf{Effect of donor placement on the magnitude of superexchange calculated using atomistic FCI along the [100] direction.} Keeping the separation of the middle donors fixed ($r_M$ = $5.431nm$), we change $r_{1,2}$ to account for any deviation in donor placement. The superexchange $\Delta E$ is maximum when $r_1$ = $r_2$. We also see that $\Delta E$ remains constant when $r_1 + r_2$ is constant -- see dashed lines.}
    \label{fig4}
\end{figure}
The results are shown in Figure \ref{fig4}. Here we can see the superexchange $\Delta E$ is maximum (11.53 GHz) for the smallest values of $r_1$ and ($r_1$ = $r_2$ = $7.6$ nm). This result arises both from minimizing the total chain separation, $R$ as well as maximizing the $r_M/r_{1,2}$ ratio -- as observed in Figure \ref{fig3}. The value of $\Delta E$ decreases exponentially if either of the outer donors is moved outwards, i.e. if either $r_1$ or $r_2$ increases. However, at the same time, we can see $\Delta E$ does not change if both of the outer donors are shifted simultaneously in the same way -- see the dashed lines where $r_1+r_2$ is constant. This result is consistent with the Schrieffer-Wolff approximation for superexchange in Eq. \ref{HSW}, where the dominating contribution to $\Delta E$ comes from $j_1j_2/2j_M$. While $j_1$ and $j_2$ change approximately exponentially with distance, their product will not change when $r_1+r_2$ is kept constant. 

Thus we can conclude that any small shifts in the location of the donors within the [100]-oriented chain can result in large changes in total $\Delta E$. However, interestingly, a simultaneous shift of both outer (or both middle) donors in the same direction would not have any impact on the superexchange (indicated by the black dashed lines). The situation is completely different for the [110] crystalline direction due to the oscillations of two-donor exchange with distance which lifts the smooth $j_1j_2/2j_M$ dependency for varied $r_1,r_2$ when $r_1+r_2$ is constant. Here $\Delta E$ will change in an oscillatory manner if one or both middle donors are shifted.

\subsection{Impact of donor nuclear spins on the superexchange}
Finally, we comment on the presence of nuclear spins in the donor system and how they can also impact superexchange and coherent electron spin manipulation. The nuclear spins of phosphorus donor atoms couple with their electron spins through the hyperfine interaction. From the perspective of the electron spin, this can be treated as a small additional magnetic field dependent on the nuclear spin polarization. In the case of an electron singlet-triplet spin qubit, localized within the middle double donor dot, this hyperfine interaction can create a magnetic field gradient that mixes singlet and triplet states \cite{Osika2022,Kranz2022}. For a 1P-1P system, the gradient is $\sim$ few Milli Tesla which gives rise to a Zeeman energy difference, $\Delta E_z$ $\sim$ $5\times10^{-4}$ meV between the donors when the two nuclear spins are oriented antiparallel ($\Uparrow \Downarrow$) and $\Delta E_z$ $\sim 0$ when the two nuclear spins are aligned in the same direction ($\Uparrow \Uparrow$). 

To create well-defined eigenstates as well as singlet and triplet states, the exchange coupling $\Delta E$ needs to dominate over $\Delta E_z$. We can consider the 4-electron system discussed in this paper essentially as two pairs of singlet-triplet qubits -- one defined between the two inner dots and one between the two outer dots. To avoid mixing of the singlet and triplet states it is desirable for both $j_M$ and $\Delta E$ to be significantly greater than $\Delta E_z$. From Figure \ref{fig1} this is satisfied when the middle donor separation $r_M$ is smaller than 10-15 nm, which includes most of the results presented in this work. The value of superexchange $\Delta E$ ultimately depends both on $r_M$ and $R$. From Figure \ref{fig3} we can see it is possible to design 4P configurations which belong to the regime where $j_1,j_2/j_M < 0.3$ and at the same time satisfy $\Delta E > \Delta E_z$. These considerations are however no longer relevant if we deterministically initialize the nuclear spins to all-parallel before any operation using a local  Nuclear Magnetic Resonance (NMR) \cite{plaSingleatomElectronSpin2012}. In this case, there is no magnetic field gradient and thus no limitations on the minimum values of $j_M$ and $\Delta E$.

\section{Conclusion}
In this work, we have presented the results of strong non-nearest neighbour exchange coupling over a long distance (20-35 nm) with a possible extension of up to 45 nm, which has not been explored in donor quantum dots before. Using an atomistic full configuration interaction technique, we have calculated the eigenvalues of a 4-electron system with high-quality atomistic basis states and showed a comparison with an effective spin Hamiltonian to determine the viability of our intensive numerical calculations. We have shown that by placing the mediator donors one atomic position inwards compared to the symmetric chain, it is possible to enable coherent control of the qubits separated in [100] and [110] directions. Similarly to previous works on direct exchange, we observe a monotonic dependence of superexchange versus distance for donors separated in [100] and oscillatory dependence in [110]. Importantly, we demonstrate that the superexchange can be modulated by an order of magnitude by purely electrostatic means using realistic experimental gate voltages. These results pave the way for the realization of fast singlet-triplet control for distanced qubits. The calculations support the experimental realization of long-distance coupling of donor qubits in silicon. 

\bibliographystyle{apsrev4-2}
\bibliography{ref}
\newpage
\appendix
\section{Convergence of multi-electron states}
\begin{figure}[htb!]
    \centering
    \includegraphics[width=0.8\textwidth]{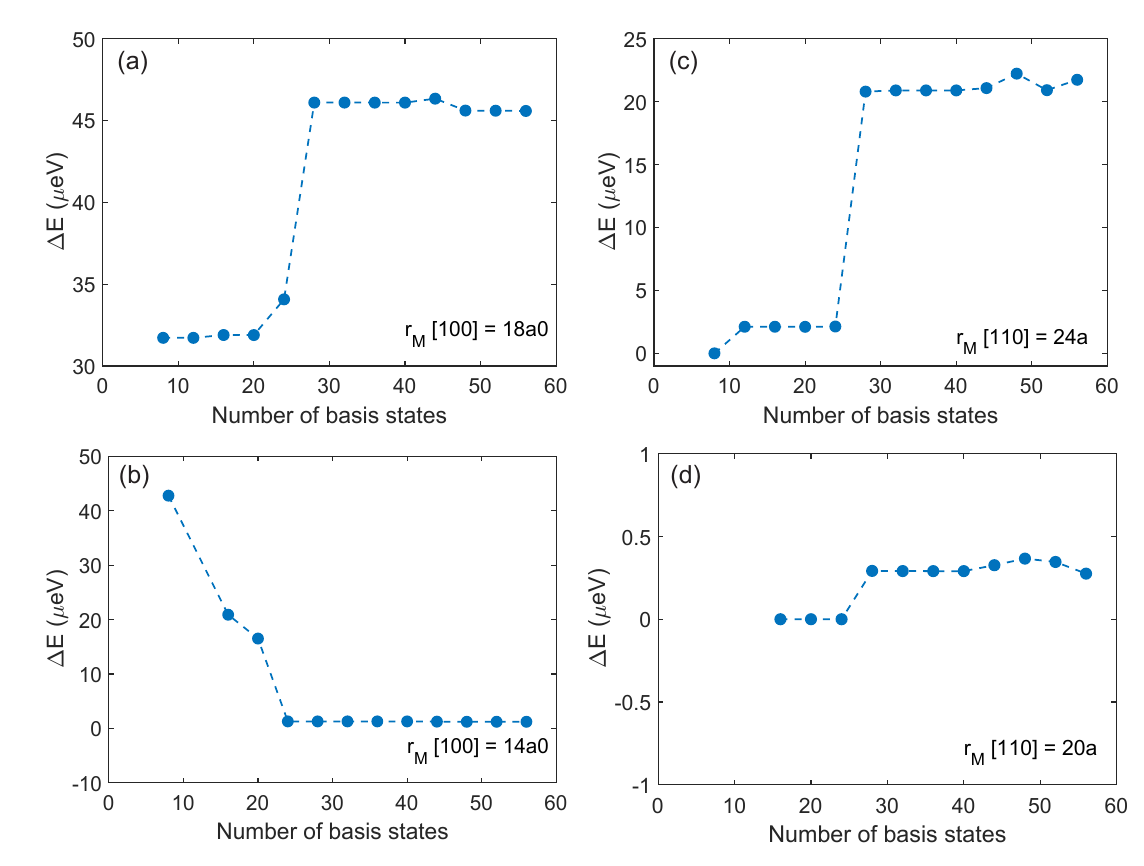}
    \caption[Convergence of superexchange as a function of a number of single-electron basis states]{\textbf{Convergence of superexchange as a function of a number of single-electron basis states.} Plots (a,c) show results for equidistant dots with $r_1=r_2=r_M=18a_0$ along the [100] direction where $a_0 = 0.5431$ nm and $r_1=r_2=r_M=24a$ along the [110] direction where $a = a_0/\sqrt{2}$, respectively. Plots (b,d) show results for non-equidistant cases $r_1=r_2=20a_0$, $r_M=14a_0$ in [100] and $r_1=r_2=26a$, $r_M=20a$ in [110], respectively.}
    \label{fig:convergence}
\end{figure}

The convergence of multi-electron states calculated using FCI depends on the number of single-electron states taken into the basis. As in this work, we investigate the effect of superexchange, i.e. the energy difference $\Delta E$ between outer-dot singlet and triplet states, $S^l$ and $T^l_0$ respectively, we use $\Delta E$ also to determine the convergence of our calculated four-electron states. We gradually increase the number of single-electron basis states in FCI calculations and observe when the changes in $\Delta E$ are negligible within some numerical tolerance. In Figure \ref{fig:convergence} we show examples of such analysis for donors separated in both [100] and [110] crystal directions. We find that in most of the cases considered in the paper, it is sufficient to use 56 basis states, i.e. 28 valley-orbital states, with two-fold spin degeneracy. In a few cases, we have used up to 72 basis states to reach convergence in the energy difference. 

\section{Analysis of four-electron eigenstates}

\begin{figure}[htbp]
    \centering
    \includegraphics[width=\columnwidth]{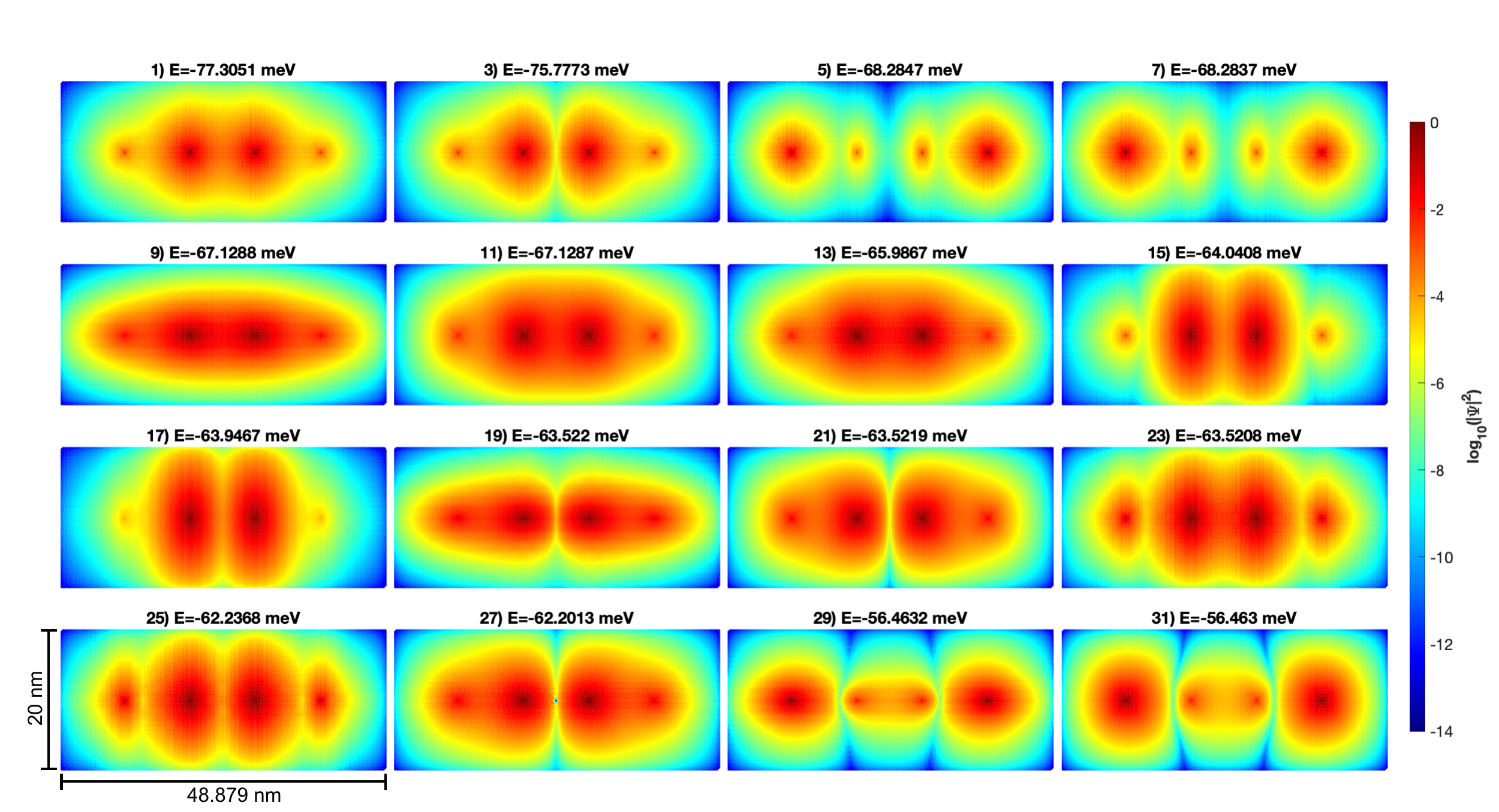}
    \caption{\textbf{Single electron basis states calculated using an atomistic tight-binding approach.} The wavefunctions of the first 16 valley-orbit molecular states are plotted here on a logarithmic scale. The titles indicate the binding energy of each of the states.}
    \label{fig:basis_states}
\end{figure}

In this section, we analyze the 4-electron states in terms of their contributions from different Slater determinants, i.e. consisting of different 1-electron atomistic tight-binding basis states. This will help us to determine the regimes of $j_1,j_2,j_M$ where the eigenstates can be confidently described in terms of singlet-triplet symmetries. First, we focus on [100] crystal direction and the equidistant case $r_1=r_2=r_M=18a_0$. In Figure \ref{fig:basis_states} we plot several of the lowest-energy single electron states. 

\begin{itemize}
    \item 
We can see that the ground and first-excited orbitals (states numbered 1 and 3 in the figure) are localized mainly in the two middle dots. This is because superposing the four donors confining potentials in TB calculations results in the total potential being slightly deeper within the middle dots than the outer ones. 
    \item 
The next two excited orbitals (states 5 and 7) are localized mainly within two outer dots. 
    \item
We consider the even-numbered states (2,4,6,8...) to be of the same orbital as odd (1,3,5,7...) but of opposite spin. 
    \item
The 4-electron ground state, labelled by us as $S_l$, has main contributions from following configurations: 14.7\% $\ket{1,3,6,8}$, 14.7\% $\ket{2,4,5,7}$, 10.7\% $\ket{1,2,5,6}$, 9.5\% $\ket{1,2,7,8}$, 8.9\% $\ket{3,4,7,8}$, 7.8\% $\ket{3,4,5,6}$, and smaller contributions from other configurations.
    \item
The first two contributions can be interpreted as inner-triplet outer-triplet states ($\ket{1,3,6,8}$ is $T_-$ in inner dots and $T_+$ in outer dots,  $\ket{2,4,5,7}$ is opposite to that). 
    \item
Other configurations are inner-singlet outer-singlet states. 
    \item
We can then see that a significant part of the ground state is described as an inner-dot triplet, which is specific for $j_{1,2}\approx j_M$ regime and has been expected by us following the above discussion on effective Hamiltonian states. 
    \item
To obtain a rough estimate of the percentage of singlet and triplet contributions within the middle dots we sum up contributions from states 
$\{\ket{1,2,i,j},\ket{3,4,i,j}\}$ 
for singlet and $\{\ket{1,3,i,j},\ket{1,4,i,j}$,
$\ket{2,3,i,j}$,
$\ket{2,4,i,j}\}$ for triplet. 
\end{itemize}
We show the results in Table \ref{tab:contrib}. As expected from the effective Hamiltonian model here both contributions are close to 50\%. The presented numbers are, however, just an estimate, not a precise evaluation of singlet and triplet admixtures in the middle dots as i) we do not sum up all the higher-order Slater determinant probabilities and ii) the orbitals 1 and 3 have small but non-zero probabilities also in the outer dots.

Next, we look at the non-equidistant case, with $r_1=r_2>r_M$ which guarantees $j_1=j_2<j_M$. In Table \ref{tab:contrib} we compare inner-singlet and inner-triplet contributions in $S_l$ for 3 values of $r_M$, i.e. $18a_0$, $16a_0$ and $14a_0$, while keeping the total four-donor separation $R$ constant at $54a_0$. We can see the inner-singlet (inner-triplet) contribution is significantly increasing (decreasing) when the middle donors are pulled closer together.

\begin{table}[htb!]
    \centering
    \begin{tabular}{|c|c|c|}
    \hline
     Separation ($a_0$) & Inner singlet contribution & Inner triplet contribution  \\ \hline
     18 & 0.4927 & 0.483  \\ \hline
     16 & 0.9209 & 0.06706 \\ \hline
     14 & 0.97904 & 0.00238 \\ \hline
\end{tabular} 
    \caption{\textbf{Approximate contributions of the inner-singlet and inner-triplet configurations in the FCI ground state of a 4-donor chain separated along the [100] direction} Here we show 3 different middle donor separations $18a_0$, $16a_0$ and $14a_0$ with a constant outer donor separation of $54a_0$. We see that the inner singlet contribution to the ground state dramatically increases from $\sim49\%$ to $\sim98\%$ as the separation decreases by $4a_0$.}
    \label{tab:contrib}
\end{table}

\end{document}